\documentclass[11pt, oneside]{article}   	
\usepackage{geometry}                		
\usepackage{graphicx}				
\usepackage{hyperref}
\usepackage{amssymb}
\usepackage{amsthm}
\usepackage{amsmath}

\theoremstyle{remark}

\newtheorem*{Q}{\bf Question}

\def\dalemb#1#2{{\vbox{\hrule height .#2pt
        \hbox{\vrule width.#2pt height#1pt \kern#1pt
                \vrule width.#2pt}
        \hrule height.#2pt}}}

\def\0{{\sst{(0)}}}
\def\1{{\sst{(1)}}}
\def\2{{\sst{(2)}}}
\def\3{{\sst{(3)}}}
\def\4{{\sst{(4)}}}
\def\5{{\sst{(5)}}}
\def\6{{\sst{(6)}}}
\def\7{{\sst{(7)}}}
\def\8{{\sst{(8)}}}

 \let\t=\tau

 \def\bd{\begin{document}} \def\ed{\end{document}}
\def\ds{\documentstyle} \let\fr=\frac \let\bl=\bigl \let\br=\bigr
\let\Br=\Bigr \let\Bl=\Bigl
\let\bm=\bibitem
\let\na=\nabla
\let\pa=\partial \let\ov=\overline
\newcommand{\be}{\begin{equation}}
\newcommand{\ee}{\end{equation}}
\def\ba{\begin{array}}
\def\ea{\end{array}}
\def\ft#1#2{{\textstyle{{\scriptstyle #1}\over {\scriptstyle #2}}}}
\def\fft#1#2{{#1 \over #2}}
\def\del{\partial}
\def\sst#1{{\scriptscriptstyle #1}}
 \def\oneone{\rlap 1\mkern4mu{\rm l}}
\def\ie{{\it i.e.\ }}
\def\via{{\it via}}
\def\semi{{\ltimes}}
\def\str{{\rm str}}
\def\Dm{{{D_{\sst{max}}}}}
\def\vac{ \left | 0 \right \rangle }
\def\kvac{ \left | k \right \rangle }

\def\sp{\; \; \;}

\def\bol{ \left | B (p^+) \right \rangle}
\def\bo1{ \left | B^0 (p^+) \right \rangle}

\def\bolt{ \left | B (p^+) \right \rangle_{\t}}

\def\boxl{ \left | B (x^-) \right \rangle}

\def\<{ \langle }
\def\>{ \rangle }

\usepackage{color}
\def\amklink#1#2{\href{http://andreimikhailov.com/math/bv/#1}{\textcolor{blue}{\bf #2}}}
\newcommand{\anote}[1]{\marginpar{\hspace{5pt}\hbox{\vline\vline\hspace{2pt}\vbox{#1}}\hspace{5pt}}}
\newcommand{\commentstarts}{\begin{centering}
\hspace{-1pt}\vrule\vrule
\begin{minipage}[t]{0.03\linewidth}
\hspace{0.025\linewidth}
\end{minipage}
\begin{minipage}[t]{0.95\linewidth}}
\newcommand{\commentends}{\end{minipage}
\end{centering}
\vspace{7pt}
}

\title{Families of gauge conditions in BV formalism }
\author{Andrei Mikhailov\footnote{Instit\'{u}to de Fi\!${}'$sica Te\'{o}rica, Universidade Estadual Paulista, S\~{a}o Paulo, Brazil}\;\;\footnote{on leave from Institute for Theoretical and Experimental Physics, Moscow, Russia} 
\and 
Albert Schwarz\footnote{Department of Mathematics, University of California, Davis, USA}}
\date{}							

\begin{document}
\maketitle
\begin{abstract} In BV formalism we can consider a Lagrangian submanifold as a gauge condition. Starting with the BV action functional we construct a closed form on the space of Lagrangian submanifolds. If the action functional is invariant with respect to some group $H$  and $\Lambda$ is an $H$-invariant family of Lagrangian submanifold then under certain conditions we construct  a form  on $\Lambda$ that descends to a closed form on $\Lambda/H.$ Integrating the latter form over a cycle in  $\Lambda/H$  we obtain numbers that can have interesting physical meaning. We show that one can get string amplitudes this way. Applying this construction to topological quantum field theories one  obtains topological invariants.
\end {abstract}
\section {Introduction}
A physical theory can be represented by various equivalent action functionals. For example, in the case of degenerate action functionals we can impose different gauge conditions. In BRST-formalism infinitesimal Q-exact variation  of action functional  leads to equivalent action functional. In BV-formalism the role of choice of gauge condition is played by the choice of Lagrangian submanifold.

As an example one can consider topological quantum field theories of Witten type, where the  action functional in BRST-formalism depends on metric, but the variation of this functional by an infinitesimal change of the metric (the energy-momentum tensor) is Q-exact. 

The first impression is that  it is sufficient to consider only one functional from  a family of physically equivalent action functionals . As was noticed in \cite {ST} this is wrong. The consideration of a family of equivalent action functionals  or family of gauge  conditions   labeled by points of (super) manifold $\Lambda$ leads to a construction  of a closed differential form $\Omega$ on $\Lambda$  (a  closed pseudodifferential form if $\Lambda $ is a supermanifold).  If our action functionals are invariant with respect to some group $H$ then the form $\Omega$  is $H$-invariant, but it does not necessarily descend to $\Lambda/H.$  Under some conditions we construct a closed $H$-equivariant form $\Omega _H$ and show that this equivariant form is homologous to a form descending to $\Omega/H$. This allows us to obtain interesting physical quantities integrating over cycles in $\Lambda/H.$ 

For example, we can start with topological quantum field theory on some manifold $\Sigma$. One can apply our results to the family of equivalent action functionals labeled by metrics on $\Sigma$. We obtain topological invariants of $\Sigma$ this way; it would be interesting to calculate them and compare with known invariants.

This machinery can be applied to string amplitudes. The worldsheet of bosonic string can be considered  as 
two-dimensional topological quantum field theory. Considering $\Lambda$ as a space of metrics and $H$ as a group 
generated by diffeomorphisms and Weyl transformations we get formulas for string 
amplitudes; for appropriate choice of Lagrangian submanifolds these formulas coincide with the standard ones.
Similar constructions work for other types of strings.

\paragraph     {Some remarks about terminology and notations.}

 We are saying "manifold" instead of "supermanifold", "group" instead of "supergroup", etc.  We understand an element of super Lie algebra as a linear combination $\sum \epsilon^AT_A$ where $T_A$ are even or odd generators of $\mathbb{Z}_2$ -graded Lie algebra and $\epsilon^A$ are even or odd elements of some Grassmann algebra; hence in our understanding an element  of super Lie algebra is always an even object
   (see \cite{SC}, \cite{SD} for the definitions of supermanifold , super Lie algebra, etc. that we are using).
   
We work in BV-formalism assuming that the BV action functionals  are defined on  odd symplectic manifold $M$ equipped with volume element (SP-manifold in terminology of \cite {Schwarz:1992nx,Schwarz:1992gs}) . In this situation the odd Laplacian $\Delta$ is defined on the space of functions on $M$. It was noticed in \cite {Khudaverdian:1999} that in the absence of the volume element the odd Laplacian is defined on semidensities; this allows the reformulation of BV-formalism for any odd symplectic manifold. In Appendix C we show how to prove our main results in this more general setting. Some basic formulas of BV-formalism are listed in Appendix A.

The space of (smooth) functions on a supermanifold $M$ is denoted  $\mbox{Fun}(M)$.This space is $\mathbb{Z}_2$-graded: $\mbox{Fun}(M)=\mbox{Fun}_{\bar{0}}(M)+\mbox{Fun}_{\bar{1}}(M).$ Functions on $\Pi TM$ (on the space of tangent bundle with reversed parity of fibers) are called pseudodifferential forms (PDF) on $M.$ (Differential forms can be considered as polynomial functions on $\Pi TM$.) Diff stands for the group of diffeomorphisms, Vect for its Lie algebra (the algebra of vector fields), 
Weyl for the group of Weyl transformations. 
As we have noticed an element of any super Lie algebra  (and hence a vector field) is considered an even object.

We use the term "canonical transformation" for a transformation of (odd) symplectic manifold preserving the symplectic form (another word for this notion is "symplectomorphism").  On a simply connected manifold infinitesimal canonical transformations can be characterized as Hamiltonian vector fields. Notice that in our terminology  the Hamiltonian on odd symplectic manifold is an odd function $B$; the first order differential operator corresponding to the Hamiltonian vector field with the Hamiltonian $B$ is expressed in terms of the odd Poisson bracket as an operator transforming a function $G$ into $\{B,G\}$; this operator is even (parity preserving). The condition $\Delta B=0$ means that the Hamiltonian vector field is volume preserving (= divergence free).

\section{Families of equivalent action functionals}\label{sec:FamiliesOfActionFunctionals}
Let us consider a  functional $S$ defined on an odd symplectic  
manifold $M$ with volume element 
and satisfying the quantum master equation $\Delta e^{S_{\rm BV}}=0.$  (Here $\Delta$
stands for the \amklink{BV-formalism/OddLaplace.html}{odd Laplacian}.) 
Then the physical quantities corresponding
to the BV action functional $S_{\rm BV}$ can be expressed as integrals $\int_LAe^{S_{\rm BV}}$ 
where $L$ is a Lagrangian submanifold of $M$ and the integral is taken with respect to the volume
element induced on this submanifold;  $A$ stands for quantum observable
(i.e. $\Delta (Ae^{S_{\rm BV}})=0$ or equivalently $\Delta A+ \{A, S_{\rm BV}\}=0$). These integrals depend only on the 
homology class of the
Lagrangian submanifold. 

Let us consider now a family of 
physically equivalent BV-action functionals $S_{\lambda}, \lambda \in \Lambda$ obeying $\{S_{\lambda},S_{\lambda}\}=0$, $\Delta S_{\lambda}=0$.  We can consider $S$ as a 
function on $\Lambda\times M$. We assume that $\Lambda$ is simply connected; then $S_{\lambda}$ being physically equivalent for 
different values of $\lambda$ is equivalent to the existence of functions $B_a$ such that:
\begin{equation}\label{s}
{\partial\over\partial\lambda^a}S_{\lambda}=\{B_a, S_{\lambda}\}
\end{equation}
for some $B_a\in \mbox{Fun}_1(M)$, $\Delta B_a=0$  (one can describe  $B_a$ as Hamiltonians of infinitesimal volume preserving canonical transformations giving equivalence of functionals  $S_{\lambda}$ for infinitesimally close $\lambda$).  The Eq. (\ref {s}) implies that $\left\{{\partial B_a\over\partial\lambda^b} - {\partial B_b\over\partial\lambda^a} + \{B_a,B_b\}\;,\;S_{\lambda}\right\} \;=\; 0$. 
We will assume  a stronger condition :
\begin{align}\label{ZeroCurvatureOnB}
 dB - {1\over 2}\{B,B\} \;=\; & 0
\\   
 \mbox{\tt\small where } & B = d\lambda^a B_a
\end{align}
Then the following PDF on $\Lambda$ is closed:
\begin{equation}
   \Omega(\lambda,d\lambda) \;=\; \int_L \exp\left(S_{\lambda} + B\right)
\end{equation}
Indeed using Eqs.  (\ref{DeltaEPhi}) and (\ref{DeltaInDarboux}) we obtain
\begin{align}
   d\Omega(\lambda,d\lambda)\;=\;&
   \int_L \left(\{B,S\} + {1\over 2}\{B,B\}\right)e^{S+B}\;=\;\int_L \Delta e^{S+B} \;=\;0
\end{align}
More generally, let us define:
\begin{align}\label {DOF}
   \Omega\langle F\rangle (\lambda,d\lambda) \;=\;& \int_L F e^{S+B}
   \\    
   \mbox{\tt\small where } & F\in \mbox{Fun}(\Lambda\times M) \mbox{ \tt\small such that }
   dF = \{B,F\}
\end{align}
Then:
\begin{equation}
   d\Omega\langle F\rangle = - \Omega \langle \Delta F + \{S,F\}\rangle
\end{equation}
Eq. (\ref{DOF}) follows from the following chain of equalities:
\begin{align}
d\Omega \;=\; & \int_L \left(\{B,F\} + {1\over 2} \{B,B\}F + \{B,S\}F\right) e^{S+B} \;=
\\   
\;=\; & \int_L \Delta\left(Fe^{S+B}\right) \;-\; \int_L (\Delta F + \{S,F\}) e^{S+B}
\end{align}
and $\int_L \Delta(\ldots)=0$.

Notice  that Eq. (\ref{s}) does {\em not} define $B_a$ unambiguously; there is a freedom to
add to $B_a$ a function $\{S,A_a\}$ where $\Delta A_a=0.$
   One can use this freedom  to obtain $B_a$ satisfying (\ref {s}) and (\ref{ZeroCurvatureOnB}). This is not always possible  
   globally, but always possible locally (in small pieces of the parameter space $\Lambda$). To check this we consider a fiber bundle over $\Lambda$ having as a fiber over a point $\lambda \in \Lambda$ the set of volume preserving canonical transformations  transforming $S_{\lambda_0}$ in $S_{\lambda}. $ (Here $\lambda_0$ is a fixed point of $\Lambda$.) A continuous (even differentiable) section of this bundle not  necessarily exists globally, but always exists locally. It exists globally, in particular, in the case when $\Lambda$ is contractible. Differentiating the section $U_{\lambda}$ we obtain infinitesimal canonical transformations $\hat B_a=\frac{\partial U}{\partial \lambda^a}U^{-1}$.  Their Hamiltonians $B_a$
   obey (\ref {s}) and (\ref{ZeroCurvatureOnB}). (This is not quite correct: the operators $\hat{B} = d\lambda^a \hat{B}_a$ obey 
$d\hat{B} - 1/2[\hat{B},\hat{B}]=0$, but their Hamiltonians $B_a$  specified {\it via} $\hat{B}_a = \{B_a,\_\}$
are defined only up to a $\lambda$-dependent constant and  (\ref{ZeroCurvatureOnB}) is true only for an appropriate 
choice of these constants; see Appendix \ref{sec:CentralExtension} for details.)

\section{Families of Lagrangian submanifolds in BV phase space}\label{sec:FamiliesOfLAG}
 We will show that one can construct  some
interesting quantities (including string amplitudes) considering {\em families}
of Lagrangian submanifolds instead of families of action functionals. 

Let us fix a connected family $\Lambda$ of simply connected Lagrangian submanifolds. In
other words we assume that $L$ depends on parameters $\lambda_1, ...,
\lambda_k,...$ 
(these parameters can be odd, but for simplicity we assume
that they are even). Let $G$ be the group of canonical transformations of $M$
(transformations preserving the odd symplectic structure), and $\bf g$ its Lie algebra. Elements of 
$\bf g$ correspond to odd functions on $M$ (Hamiltonians).

\paragraph     {Tentative definition of the closed form $\Omega$}
We want to define a closed pseudo-differential form $\Omega$ on the space $\rm LAG$ of all simply-connected 
Lagrangian submanifolds:
\begin{equation}
\Omega\;\in\; \mbox{Fun}(\Pi\;T\;\;{\rm LAG})
\end{equation}
Roughly speaking, the value of $\Omega$ at a point $v \in \Pi\;T\;{\rm LAG}$ is computed as follows. 
Notice that $v$ corresponds to a pair $(L,\sigma)$ where $L\in {\rm LAG}$ and $\sigma\in \mbox{Fun}(L)$ is an odd function 
on $L$ describing the tangent vector\footnote{
As in Classical Mechanics, a function on Lagrangian manifold $L$ specifies a tangent vector to LAG (an infinitesimal deformation of $L$). In our case the symplectic form is odd, hence the correspondence is parity reversing . These functions are called ``infinitesimal gauge fermions''. We have assumed that $L$ is simply-connected, in this case the map of functions to infinitesimal deformations is surjective and its kernel consists of constant functions .}  The variation of $L$ can be described by infinitesimal canonical transformation; one can say that $\sigma$ is a restriction to $L$ of the Hamiltonian of this transformation. (Notice that the canonical transformation is not unique, but the restriction of its Hamiltonian to $L$ is well defined up to a constant summand.) In other words, for any vector field $v$ inducing a tangent 
vector to ${\rm LAG}$ at $L$ we have: 
\begin{equation}
   d\sigma \;=\; - \left.\left(\iota_v \omega\right)\right|_L.
\end{equation}
The function $\sigma$ depends on $v\in \Pi\;T_L\;{\rm LAG}$ (on odd tangent vector to ${\rm LAG}$ at $L$) linearly, hence it can be considered as a one-form on  ${\rm LAG}.$ 

By definition:
\begin{equation}
  \Omega(L,v) = \int_L e^{S_{\rm BV} + \sigma}
  \label{omega-in-BV-formalism}
\end{equation}
More generally, for every function $F$ on $M$ we define:
\begin{equation}
   \Omega\langle F\rangle(L,v)\;=\;\int_LFe^{S_{\rm BV}+\sigma}
\end{equation}
As a complication, the one-form $\sigma$ is defined only up to a constant:
\begin{equation}\label{ShiftOfSigma}
\sigma \;\mapsto\; \sigma + \mbox{const}
\end{equation}
Therefore the definition of $\Omega$ by Eq. (\ref{omega-in-BV-formalism}) is strictly speaking ambiguous. We will prove that it is always possible 
to resolve this ambiguity in such a way, that  the form $\Omega$ is closed. Moreover, 
\begin{equation}
\label {ofd}
   d\Omega\langle F\rangle = - \Omega \langle \Delta F + \{S,F\}\rangle
\end{equation}
It is enough to prove this formula for restriction to any finite-dimensional
submanifold $\Lambda\subset {\rm LAG}$ ({\it i.e.} a family of Lagrangian submanifolds). Let us parameterize
$\Lambda$ by coordinates $\lambda^1,\ldots,\lambda^n$. This means that we have a family of Lagrangian submanifolds $(L(\lambda))$.

Let us find a family of volume preserving canonical transformations $g(\lambda)$ such that:
\begin{align}
   L(\lambda) \;=\; & g(\lambda)L_0
\end{align}
(locally this is always possible).
The introduction of such $g(\lambda)$ is essentially a trick. It does not participate in any way
in the {\em definition} of $\Omega$; we will use it just to compute $d\Omega$. Using $g(\lambda)$
we can construct a family of physically equivalent action functionals $S_{\lambda}$ obeying 
 $$\int _{L_0}e^{S_{\lambda}}=\int _{L_{\lambda}}e^S.$$
Here $S_{\lambda}$  is obtained from $S$ by means of the transformation $g_{\lambda}$. It is easy to check
that the form $\Omega$ introduced in present Section  coincides with the form constructed in the Section 2 for the family $S_{\lambda}$   and denoted by the same symbol; hence it is closed. (The second summand in the definition of $\Omega$ in section 2 is a Hamiltonian $\cal H$ of the infinitesimal canonical transformation governing the variation of $S_{\lambda}$. 
The  Hamiltonian governing the variation of $L_{\lambda}$   enters  the definition of $\Omega$ in present section. These two Hamiltonians coincide up to a constant; resolving the ambiguity  in the definition of second Hamiltonian in appropriate way we can say that the Hamiltonians coincide.)

If we know the precise definition of $\Omega$ we can give also a precise definition of  $\Omega\langle F\rangle.$ The formula (\ref {ofd}) follows from (\ref{DOF}).

\vspace{10pt}
\noindent
A more formal proof of the results of this section is given in  Appendix \ref{sec:CentralExtension}.

\section{Gauge symmetries}\label{sec:GaugeSymmetries}
\paragraph     {Form $\Omega$ is not necessarily base with respect to gauge symmetries}
We assume that the action functional $S$, the observable $A$, the volume element on
$M$, and the family $\Lambda$ are invariant 
under a subgroup
$H\subset G$ ( or Lie algebra ${\bf h}\subset \bf g$). \footnote {Notice, that $H$ is not necessarily the full group of automorphisms. In string worldsheet theory, typically $H$ is the
group of diffeomorphisms.}  We denote by $\bf \hat h$  the set of Hamiltonians of  elements of $\bf h$; then the $\bf h$-invariance of $S, A$ and volume element 
means that for every $h\in \bf\hat h$ we have $\{S,h\}=0, \{A,h\} =0$ and $\Delta h=0.$   
(It is enough to impose a weaker requirement:
\begin{equation}\label{DeltaHisZero}
   \Delta h + \{S_{\rm BV},h\} \;=\; 0,
\end{equation}
see \cite{Schwarz:1993xp}.)
It follows from these assumptions that the form $\Omega$ is also 
$H$-invariant (or ${\bf h}$-invariant).  In general the form $\Omega$
is {\em not horizontal}, and therefore 
\amklink{omega/Descent\_To\_Double\_Coset.html}{does not descend to $\Lambda/H$}.
However, in some important cases, in particular in string theory, the  form $\Omega$ does descend to  $\Lambda/H$ for appropriate choice of the family of Lagrangian submanifolds.

We will now construct a modified form $\Omega$ which is base. 

Under the assumptions of previous section , let us make the following 
\amklink{introduction/Proof_of_equivariance.html}{additional assumption}. 
Suppose 
that there exists a map $\Phi\;:\;{\bf \hat h}\to \mbox{Fun}(M)$ such that every  Hamiltonian $h\in{ \bf \hat h}$ satisfies:
\begin{equation}
\label{ph}
h=\{S_{\rm BV},\Phi(h)\}+\Delta \Phi(h)+ {1\over 2}\{\Phi(h),\Phi (h)\}
\end{equation}
(Notice that the Hamiltonian $h$ is odd, but $\Phi (h)$ is even.) We will also require that $\Phi$ satisfies
the following ``equivariance'' property. For any two elements $h\in\hat{\bf h}$ and ${\tilde h}\in\hat{\bf h}$:  
\begin{equation}\label{EquivarianceOfPhi}
\{h,\Phi(\tilde h)\}\;=\;\Phi(\{h,\tilde h\})
\end{equation}
Let us suppose that the action of $\bf {h}$ on $\Lambda$  comes  from  a free action of the corresponding  
Lie group $ H$  (this Lie group is not necessarily connected). Then  
\amklink{omega/Descent\_To\_Double\_Coset.html\#(part.\_.Modified\_.P.D.F)}{we can construct}
closed form $\Omega_{H}$, which descends to $\Lambda/H$. (In other words this is a base form, {\it i.e.} $H$-invariant and $H$-horizontal form.)

Technically, we use the  
\amklink{equivariant-cohomology/index.html}{formalism of equivariant cohomology}.
The conditions we impose on the map $\Phi$ allow us to prove that 
 the  form \begin{equation}\label{EquivariantOmega}
   \Omega^{\tt C}_{H}(\lambda, d\lambda,h) = \int_{L_{\lambda}} e^{S_ {\rm BV}+ \sigma + \Phi (h)}
\end{equation}
represents a class of $H$-equivariant cohomology of $\Lambda$ in the 
\amklink{equivariant-cohomology/Equivariant\_Cohomology.html}{Cartan model}. (We consider here $\sigma$ as a one-form on $\Lambda.$)

Recall that in this model an  equivariant cohomology class is represented by a differential 
form depending on an element of $\bf h$ and belonging to the kernel of Cartan differential $d-\iota _{h}$ 
where  $h\in \bf h$. (The dependence of  $\bf h$ should agree with the action of the group $H.$) We  modify the definition allowing pseudodifferential forms instead of differential forms. We do not 
impose the condition of polynomial dependence of  $\bf h$.

\noindent
The proof of the fact that Eq.  (\ref{EquivariantOmega}) is equivariantly closed uses (\ref {ofd}) and
\amklink{omega/As\_Intertwiner.html}{the relation}
 \begin{equation}
\label{h}
\iota _{r}\Omega \langle F\rangle = \Omega \langle RF\rangle
\end{equation}
where $r\in \bf g$ and $R$ stands for the corresponding Hamiltonian. This formula immediately follows from:
\begin{equation}\label{ActionOfIotaOnSigma}
\iota_r \sigma = R|_L
\end{equation}
which is essentially the definition of $\sigma$.

In the case when $H$ is a conventional group  the Poisson bracket corresponds to usual commutator hence $\{h,h\}=0$; combining this with Eq. (\ref{EquivarianceOfPhi}) we get:
\begin{equation}
   \{h,\Phi(h)\} = 0
\end{equation}
(this also can be derived just from Eqs. (\ref{DeltaHisZero}) and (\ref{ph})).

{\bf From Cartan to base}

If the action of $H$ on $\Lambda$ is free the $H$-equivariant cohomology is isomorphic to the cohomology of $\Lambda/H.$
An explicit formula for a base form belonging to the same class of equivariant cohomology as   $ \Omega^{\tt C}_{\bf  h}$ 
can be written  
\amklink{equivariant-cohomology/Direct\_Computation.html\#(elem.\_.Def.Underline.Alpha)}{as follows}. 
We need to choose a connection $\theta$ on $\Lambda$ (the cohomology class of the resulting base form will not
depend on the choice of $\theta$). Then we have to replace $\sigma$ with the horizontal projection of $\sigma$,
and substitute the curvature $f=d\theta - {1\over 2}\theta^2$ for $h$ (see \cite{Meinrenken} for a review): 
\begin{equation}\label{OmegaBase}
   \Omega^{\tt base} = \int_{L_{\lambda}} \exp\left[
      S_{\rm BV} + (\sigma - \iota(\theta)\sigma) + \left(d\theta -{1\over 2}\theta^2\right)\Phi
   \right]
\end{equation}
The second term $\sigma -\iota(\theta)\sigma$ is the horizontal projection of $\theta$. 
The third term $\left(d\theta - {1\over 2}\theta^2\right)\Phi$ should be understood as follows. Consider the
curvature $d\theta - {1\over 2}\theta^2$ of the connection in the fiber bundle $\Lambda \rightarrow \Lambda/H$; this is an
$H$-equivariant $\bf h$-valued 2-form on $\Lambda$. Composing it with our map $\Phi$ we get a two-form with values in $\mbox{Fun}(M)$,
which is denoted $\left(d\theta -{1\over 2}\theta^2\right)\Phi$ in Eq. (\ref{OmegaBase}). 

\vskip .3in   

The  considerations above are rigorous in finite-dimensional case, however, we will use them in infinite-dimensional case where they can be justified in the framework of perturbation theory. 
Notice in the case when the dimension is infinite one should impose some additional conditions.  In particular, the quadratic part of the BV action functional restricted to the Lagrangian submanifold should be non-degenerate. This condition (non-degeneracy condition) is necessary to have well defined  perturbation theory. It is not needed in finite-dimensional case when the integral has a definition independent of the perturbation theory and the integral of degenerate functional makes sense. The situation with the completeness condition is similar: it is necessary only in infinite-dimensional case.

 The odd Laplacian $\Delta$ is ill-defined  in the infinite-dimensional case unless we are working in the framework of perturbation theory when we can apply the methods of \cite {COS} or \cite {Cattaneo:2015vsa}. However  the equation $\Delta S=0$ 
does make sense; it just means that the nilpotent vector field  $Q$  
corresponding to the first order differential operator transforming a function $f$ into $\{f, S\}$ 
is volume preserving. (There exist standard ways to check that an operator in infinite-dimensional space is volume preserving; for example a method based on the calculation of Seeley coefficients is explained in \cite {SCHP}.)  Replacing $S$ by $\exp\left[\frac{S}{\hbar}\right]$  we can write the  quantum master equation 
$\Delta e^\frac{S}{\hbar}=0$ as $\{S,S\}+\hbar \Delta S=0$; in infinite-dimensional case we assume that both summands 
vanish: $\{S,S\}=0$ (classical classical master equation) and $\Delta S=0.$
Similarly, we assume that in (\ref {ph}) $\Delta \Phi=0.$  In infinite-dimensional case we require that a quantum observable $A$ satisfies the equations $\Delta A=0$ and $\{A,S\}=0.$

\section{From BRST to BV}\label{sec:FromBRSTtoBV}
Let us suppose that we have a functional $S(\psi)$ with an odd
symmetry $Q_{\rm BRST}$  (BRST symmetry) that is nilpotent {\em off-shell} ({\it i.e.} nilpotent without using the equations of motion). Then  we can construct an odd symplectic manifold adding antifields $\psi^*$  and solution to the  classical Master Equation  given by the formula 
\begin{equation}
\label{bv}
 S_{\rm BV} =  S(\psi)+ (Q_{\rm BRST}\psi^i) \psi_i^{\star}
\end{equation}
In the case when  $Q_{\rm BRST}$ is volume preserving (divergence-free)  $S_{\rm BV}$ obeys also  quantum 
master equation $\Delta S_{\rm BV}=0.$ This statement is rigorous in finite-dimensional situation; 
it remains true also in the infinite-dimensional case.

A special case of this construction  comes from  the ``standard'' BRST formalism. It works for gauge theories as
Yang-Mills/QCD or Chern-Simons, and also for the \amklink{bosonic-string/index.html}{bosonic string worldsheet theory}
and the \amklink{Heterotic-RNS/index.html}{RNS superstring} .

One starts from the ``classical action'' $S_{\rm cl}(\varphi)$, which is invariant  with respect  to group 
$H$, hence with respect to its  Lie algebra $\cal H$ with generators $T_A$ ("gauge symmetry").
Then one introduces  additional variables $c^A$ (``the ghosts'') with the quantum numbers of the symmetry
parameter, but opposite statistics. 

The nilpotent symmetry $Q$ is 
\amklink{BRST-formalism/Construction.html}{defined by the following  formulas}: 
\begin{equation}
Q_{\rm BRST}\varphi^i= T^i_Ac^A\,,\;
Q_{\rm BRST} c^A = {1\over 2} f^A_{BC}c^B c^C
\end{equation}
where $f^A_{BC}$ are structure constants of the Lie algebra $\cal H.$ To continue
\amklink{BRST-formalism/BV\_from\_BRST.html}{from BRST to BV}, we define an odd 
symplectic manifold adding to $\varphi^i, c^A$ their antifields $\varphi _i^*, c_A^*$ having opposite parity
(geometrically this means that we consider cotangent bundle with reversed parity of 
fibers). Here $\varphi^i$ is the  collective notation for the ``old fields''.
In such a situation, a solution of the
classical Master Equation (a special case of (\ref {bv})) can be written in the form:
\begin{equation}
\label {b}
S_{\rm BV} =  S_{\rm cl}(\varphi) +  {1\over 2} f^A_{BC}c^B c^C c_A^{*} + T^i_Ac^A \varphi_i^{\star}
\;=\;
S_{\rm cl}(\varphi) + (Q_{\rm BRST}c^A)c_A^{\star} + (Q_{\rm BRST}\varphi^i)\varphi_i^{\star}
\end{equation}
Our goal will be to solve the Eq. (\ref {ph}) for BV action functional (\ref {b}).   
Notice that this action functional is invariant with respect to the action of the group $H$ and its Lie algebra $\cal H$; the hamiltonian of the element $\xi=\xi^AT_A\in \cal H$  has the form $h= T^i_A\xi^A \varphi_i^{\star} + [\xi,c]^Ac_A^{\star}.$

  {\it There 
exists a solution mapping this Hamiltonian into 
$\Phi (h)= \xi^A c^{\star}_A$; it satisfies the conditions 
$\{\Phi(h),\Phi (h)\}=\Delta {\Phi(h)}=0$. }

To check  (\ref {ph}) it is sufficient to notice that
\begin{equation}
\{\,S_{\rm BV}, \xi^Ac^{\star}_A \,\} = T^i_A\xi^A \varphi_i^{\star} + [\xi,c]^Ac_A^{\star}
\end{equation}
A solution of (\ref {ph}) should obey (\ref{EquivarianceOfPhi}). To verify  this condition
we notice that $\{T^i_A\xi^A \varphi_i^{\star} + [\xi,c]^Ac_A^{\star},{\tilde \xi}^Ac_A^{\star}\}=f^A_{BC}\xi^B{\tilde \xi}^Cc^{\star}_A =[\xi,{\tilde \xi}]^Ac^{\star}_A .$

In Section \ref{sec:BosonicString} we will illustrate these calculations in the particular case of bosonic string
worldsheet theory, where $H$ is the group of diffeomorphisms.

\paragraph     {Comment about antifields}
If $\phi$ is a scalar field, we will consider $\phi^{\star}$ a density ({\it i.e.} a volume form, or 
an area form in the two-dimensional case). This is very natural:
\begin{itemize}
   \item The odd symplectic form is given by the integral of the density
      $(-1)^{\bar{\phi}}\delta \phi\wedge \delta \phi^{\star}$, {\it i.e.} $\omega = \int (-1)^{\bar{\phi}}\delta\phi \wedge\delta \phi^{\star}$
   \item A local infinitesimal field redefinition $\phi \mapsto \phi + \varepsilon V(\phi)$ is generated
      by the odd Hamiltonian $\int V(\phi)\phi^{\star}$ (in order for this integral
      to make sense, $\phi^{\star}$ should be a density).
\end{itemize}
In the same sense, we actually think of the ``variational derivative'' ${\delta\over\delta \phi}$ as a
density; it is ``generated by'' $\phi^{\star}$ in terms of odd Poisson bracket.

\section{Topological quantum field theories. Bosonic strings}\label{sec:BosonicString}
In BRST formalism  a topological quantum field theory is defined by 
a family of action functionals depending on riemannian metric  on some manifold $X$ and 
satisfying the condition that the variation of the action functional by infinitesimal variation 
of the metric is BRST exact (topological quantum field theories of Witten type). 
In BV formalism we should have solutions to the master equation $\{S,S\}=0$ depending on 
riemannian metric and obeying $dS=\{b,S\}$ where $d$ is the de Rham differential on the 
space $\rm MET$ of all metrics and $b$ is a 1-form on this space. (If $V$ is a vector field on the space 
of metrics we can write $dS/dV=\{b(V), S\}$.)  Alternatively we can assume that the solution 
to the master equation is fixed, but the Lagrangian submanifold depends on the choice of metric.

We can construct an $n$-form $\Omega_n$ on $\rm MET$ integrating $b(V_1) ...b(V_n)e^S$ over some Lagrangian 
submanifold $L$ in the space of fields. Summing the forms $\Omega_n$ we can get an inhomogeneous 

closed form $\Omega$ that can be obtained by integrating $e^{S+b}$ over $L$. Under certain conditions 
(see Section \ref{sec:FamiliesOfActionFunctionals}) one can
prove that this form is closed and descends to the quotient space of $\rm MET$ with respect to the 
action of the group $\rm Diff$ of diffeomorphisms of $X$.  We obtain a closed form on the quotient $\rm MET/Diff$; integrating this form over a cycle we can get new invariants.  In particular, applying these ideas to Chern-Simons theory one obtains invariants constructed by Kontsevich \cite {KON}; see \cite {ST} for detail.
(Another construction of these invariants was given in \cite{WIT}.)

In the rest of this Section we will outline  applications of these ideas to string
perturbation theory. The target of string theory can be regarded as two-dimensional topological quantum field theory; the above considerations can be applied to this TQFT. We will show that  string amplitudes  are particular cases of  new invariants we have mentioned. Instead of formalism of families of equivalent action functionals we will use more flexible formalism of families of Lagrangian submanifolds.

\vspace{10pt}
\noindent

\paragraph     {Bosonic string. Master action in terms of world sheet metric}

The construction outlined in Section \ref{sec:FromBRSTtoBV} works for both 
\amklink{bosonic-string/index.html}{bosonic string}
and 
\amklink{Heterotic-RNS/index.html}{RNS superstring}. 

Let us consider bosonic string. For definiteness we are writing formulas for bosonic string in flat space. (To avoid anomalies we should assume that we work in the dimension $26.$)
We start with the action functional 
\begin{equation}
\label {sg} 
   S_{\rm mat}[g,x] = {1\over 2}\int 
   \sqrt{g}g^{\alpha\beta}\;\partial_{\alpha}x^m \partial_{\beta}x^md^2\xi
\end{equation}
We integrate here over a compact surface of genus $h$ with metric $g_{\alpha \beta}$. We always assume that $h>1.$ The subindex $\rm mat$ stands for ``matter'', although this action also involves the dynamical metric $g_{\alpha\beta}$. 
This functional is invariant with respect to diffeomorphisms and Weyl transformations $g'_{\alpha \beta}=e^{\phi}g_{\alpha \beta}$; 
hence we can construct a BV action functional introducing  diffeomorphism ghosts 
$c$,  Weyl ghosts $\zeta$ and antifields to $g_{\alpha\beta}$ ,  $x^m$ and ghosts.\footnote {BV formalism was previously applied to bosonic string in \cite{Craps:2005wk}.}
Following the
general scheme outlined in Section \ref{sec:FromBRSTtoBV} we obtain:
\begin{align}
S_{\rm BV} \;=\;& S_{\rm mat}[g,x] \;+
\label{BosonicMasterActionUsingMetric}
\\   
& + \int \Big(
   ({\cal L}_cg)_{\alpha\beta}g^{\star\alpha\beta}
   + \zeta g_{\alpha\beta}g^{\star\alpha\beta}
   + ((c^{\alpha}\partial_{\alpha})x^m)x_m^{\star} + {1\over 2}[c,c]^{\alpha}c^{\star}_{\alpha}
   + ({\cal L}_c\zeta)\zeta^{\star}
   \Big)
\nonumber
\end{align}
Here ${\cal L}_c$ is the Lie derivative along the vector field $c^{\alpha}\partial_{\alpha}$. 

We now choose the Lagrangian submanifold in the following way:
\begin{align}
\label{ls}
 g_{\alpha\beta}\;=\; & g_{\alpha\beta}^{(0)} \;,\;  x^{\star} = c^{\star} = \zeta^{\star} = 0
\\   
& 
\mbox{\tt\small where }\;\;g_{\alpha\beta}^{(0)}\mbox{ \tt\small is a fixed metric}
\end{align}
The family (\ref{ls}) of Lagrangian submanifolds is closed under the action of diffeomorphisms. 
On  Lagrangian submanifold (\ref{ls}) the action is quadratic and the form  $\Omega$ is equal to:\footnote{ We denote the de Rham differential on the infinite-dimensional space of metrics by $\delta$ instead of $d$.}
\begin{align}\label{StandardOmegaBosonicString}
   & \Omega(g^{(0)},\delta g^{(0)}) = \int [Dx\, Dg^{\star}\, Dc\, D\zeta]
   \exp\left( S_{\rm BV} + \int \delta g^{(0)}_{\alpha\beta}g^{\star\alpha\beta}\right)=\\
   \;=\;&
   \int [Dx\, Dg^{\star}\, Dc\, D\zeta]
   \exp\left( S_{\rm mat}+\int \left(({\cal L}_cg)_{\alpha\beta}g^{\star\alpha\beta}
   + \zeta t
   + \delta g^{(0)}_{\alpha\beta} g^{\star\alpha\beta}\right)\right) = \\
\;=\;&\int [Dx\, Db\, Dc\,]
\exp\left( S_{\rm mat}+\int \left((\nabla_{\alpha}c_{\beta}+\nabla_{\beta}c_{\alpha})b^{\alpha \beta} + 
\delta g^{(0)}_{\alpha\beta}b^{\alpha\beta}\right)\right).
\end{align}
We  introduced the notation $t= g_{\alpha\beta}g^{\star\alpha\beta},$ $b^{\alpha\beta} =$ traceless part of $ g^{\star\alpha\beta}, $ i.e.  
$g^{\star\alpha\beta}= b^{\alpha\beta}+\frac{1}{2}t g^{\alpha\beta}$. In the transition to the last line we integrated over $\zeta$ and $t.$

\paragraph     {Non-degeneracy}  The exponential in (\ref{StandardOmegaBosonicString}) is non-degenerate. (The restriction of $S_{\rm BV}$ to the
Lagrangian submanifold of Eq. (\ref{ls}) is  
\amklink{BRST-formalism/Family\_of\_Lagrangian\_submanifolds.html\#\%28part.\_.Non-degeneracy\%29}{non-degenerate}
modulo a finite-dimensional space of zero modes of $b^{\alpha\beta}$. This finite-dimensional degeneracy is removed by the second term in the exponential of (\ref{StandardOmegaBosonicString}).)

\paragraph     {Symmetries}
The form $\Omega$ is invariant with respect to diffeomorphisms; moreover on the family (\ref{ls}) it is a base form, because 
for any worldsheet vector field $\xi$:
\begin{align}
   &
\iota_{\xi}\Omega=   \int [Dx\, Dg^{\star}\, Dc\, D\zeta]  \left(
      \int d^2z ({\cal L}_{\xi} g_{\alpha\beta})g^{\star\alpha\beta} 
   \right)
   \exp\left( S_{\rm BV} + \int \delta g^{(0)}_{\alpha\beta}\,g^{\star\alpha\beta}\right) \;=
   \\    
   =\;
   &
   \int [Dx\, Dg^{\star}\, Dc\, D\zeta]  
   \int d^2z \;\xi^{\alpha}{\partial\over\partial c^{\alpha}}\;
   \exp\left( S_{\rm BV} + \int \delta g^{(0)}_{\alpha\beta}\,g^{\star\alpha\beta}\right) 
   \;=\;0
\end{align}
To check that the last line is zero we notice that the derivative with respect to $c_{\alpha}$ under the sign of two-dimensional integral can be replaced be variational derivative under the sign of  infinite-dimensional integral.

Let us study the behavior of this form with respect to Weyl transformations $g'_{\alpha \beta}=e^{\phi}g_{\alpha \beta}$. The $0$-th component $\Omega ^0$ of inhomogeneous form $\Omega$ can be regarded as a partition function of  conformal
 field theory. The variation of partition function by infinitesimal Weyl transformation is governed by trace anomaly $\delta Z/\delta \phi= (-\frac{cR}{12}+const)Z$ where $c$ stands for the central charge and $R$ denotes the curvature of the  worldsheet . In our case  the central charge vanishes (we are working in critical dimension $d=26$; in general the central charge is equal  to $d-26$).  We see that $\Omega^0 $ does not change by Weyl transformations. The $k$-th component of the form $\Omega$ can be expressed in terms of correlation functions of the same conformal theory.  The behavior of correlation functions by Weyl transformations is governed by conformal  dimensions $\Delta_i$ of fields $\Psi_i$:
\begin{equation}
\label {cor}
 <\Psi'_1(\xi_1)...\Psi'_k(\xi_k)>_{g'}=e^{-\sum \Delta_i\Psi_i(\xi_i)}<\Psi_1(\xi_1)...\Psi_k(\xi_k)>_g
 \end{equation}
(\cite {ZZ}, formula (13,50)).
To check the Weyl invariance of $\Omega$ we notice that the dimension of $b^{\alpha \beta}$  is $2$ (it coincides with conformal dimension) and the dimension of $g_{\alpha\beta}$ is $-2$.

We have proved that 
in critical theory $\Omega$ is Weyl invariant. Moreover, it descends not only to  ${\rm MET}/{\rm Diff}$, but also to  $\rm MET/Diff\rtimes Weyl,$ that can be identified with the moduli space of  complex structures on a compact surface of genus $h.$  (A formal proof of the fact that $\Omega$ is a base form for the Weyl group repeats the proof of similar statement for Diff.)  We can get the partition function of bosonic string integrating the form over this moduli space. (Notice that we are working with inhomogeneous forms, but the integration singles out one component of this form.)

We can solve Eq. (\ref{ph}) using 
\amklink{omega/Case_Of_Standard_BRST.html}{the general considerations} of Sec \ref{sec:FromBRSTtoBV}. 
Namely, we should take a map sending a worldsheet vector field $\xi^{\alpha}(z,\bar{z})$ plus 
infinitesimal Weyl transformation $\varphi(z,\bar{z})$ to:
\begin{equation}\label{DefPhi}
   \Phi(\xi,\varphi) = \int \xi^{\alpha} c^{\star}_{\alpha}   + \varphi\zeta^{\star} 
\end{equation}
Then the functional $\{S,\Phi(\xi,\varphi)\}$ can be considered as a Hamiltonian of  infinitesimal transformation of 
fields corresponding to the vector field $\xi$ and Weyl factor $\varphi$.  This means Eq. (\ref {DefPhi}) defines a solution of 
Eq. (\ref{ph}) for the Lie algebra of the group ${\rm Diff}\rtimes\rm Weyl $ acting on the space of fields. This allows us to construct an equivariant form
\begin{equation}\label{OmegaCartanBosonicString}
   \Omega^{\tt C}_{ Lie({\rm Diff}\rtimes\rm Weyl )}(\xi,\varphi) = 
   \int_{gL} \exp\left(
      S_{\rm BV} + \sigma + 
      \int  \xi^{\alpha}c^{\star}_{\alpha} +  \varphi\zeta^{\star} 
   \right)
\end{equation}
We can then construct the \amklink{omega/Base\_Form.html}{corresponding base form} which descends to
$\Lambda/({\rm Diff}\rtimes {\rm Weyl}).$ On the standard family of Lagrangian submanifolds given by Eq. (\ref{ls})
$c^{\star}=\zeta^{\star}=0$. Therefore $\Omega^{\tt C}_{Lie({\rm Diff}\rtimes\rm Weyl ) }(\xi,\varphi)$ becomes essentially  $\Omega$ of Eq. (\ref{StandardOmegaBosonicString})

{\bf Singular metrics.}
 Notice that in the action functional (\ref{sg}) we can allow slightly singular metrics.
We say that the worldsheet metric on  a surface of genus $h$ is slightly singular if  on 
some real curves one of the eigenvalues of the metric $g_{\alpha\beta}$ vanishes and another eigenvalue remains positive.
More precisely we suppose that $g= \det g_{\alpha\beta}$  vanishes on a family of closed real curves and in the 
neighborhood of one of these curves it takes the form  $\rho^2\sigma$ where $\rho=0$ is the equation of the curve 
and $\sigma$ is a positive function.\footnote { The simplest example of this picture is a cylinder with coordinates $(\rho, \phi)$ and metric $ds^2= d\rho^2+\rho^2d\phi^2.$ Here $-a<\rho<a, 0\leq \phi<2\pi.$} It is easy to check that under these conditions
the action functional (\ref {sg}) is finite if we make an additional assumption that $x^m$ is 
constant on every closed curve where the metric is singular. The formulas for BV action (\ref {BosonicMasterActionUsingMetric}) 
and Lagrangian submanifold (\ref {ls}) can be applied to slightly singular metrics. We obtain a family 
of Lagrangian manifolds labelled by these metrics. Factorizing the topological space  $\Lambda$ of 
slightly singular metrics with respect to diffeomorphisms and  Weyl transformations we obtain the 
space $\Lambda/{\rm Diff}\rtimes {\rm Weyl}.$  Points of this space can be identified with  complex curves having 
simplest singularities (nodes). (Every closed curve  where the metric is singular should be 
contracted to a point; the metric specifies a complex structure in the complement to these points.) 
A part of this space that consists of  stable curves (curves having only finite number of automorphisms) 
can be identified with Deligne-Mumford compactification of the moduli space of algebraic curves of 
genus $h$. This is a good topological space (an orbifold). The remaining part is a "bad" (non-separable) 
space, but it does not play any role (a heuristic explanation of this fact is the remark that its 
contribution to the partition function  is suppressed by the infinite volume 
of the automorphism group).   The form of Eq. (\ref{StandardOmegaBosonicString}) descends to Deligne-Mumford space
 as a form having some singularities. To obtain physical quantities we should integrate the form over a cycle in Deligne-Mumford space; to obtain the partition function we should integrate over the fundamental cycle. (Of course, this is only a formal calculation-due to the tachyon in the spectrum of bosonic string the integral is divergent.)

{\bf Master equation in terms of complex structures.} 
A worldsheet complex structure can be specified by a field of linear  operators $I$ 
acting on tangent spaces and obeying $I^2=-1.$ Another way to specify a complex  
structure is to fix a  complex vector field $e$  such that the complex conjugate vector 
field $\bar e$ together with $e$ specifies a basis of complexified tangent space. (To relate 
these descriptions we define $e$ as the eigenvector of $I$ having eigenvalue $i.$) Notice
that $e$ is only defined up to multiplication: $e\sim ue$, where $u$ is a complex function
on the worldsheet. 
  
Due to Weyl invariance one can express the functional (\ref {sg}) in terms of complex 
structures.  We obtain  the following functional:
\begin{align}
   S_{\rm mat}[I,x]\;=\;&\int e^{\alpha}\partial_{\alpha} x \; \bar{e}^{\beta}\partial_{\beta} x d\mu
\end{align}
where the measure $\mu$ on the worldsheet is specified by the condition the vectors $e,\bar{e}$ span a parallelogram of measure $1$ in tangent space. The functional
 is invariant with respect to diffeomorphisms. We can now follow the standard
procedure by first  
\amklink{BRST-formalism/Construction.html}{introducing the diffeomorphism ghosts} $c$ 
(BRST formalism) and then 
\amklink{BRST-formalism/BV_from_BRST.html}{adding antifields}. The result is the 
Master Action of the following form:
\begin{align}
   \label{BosonicMasterAction}
   S_{\rm BV} = \;& S_{\rm mat}[I,x] 
   + \int \Big(
   ({\cal L}_cI)_{\alpha}^{\beta}I^{\star}{}^{\alpha}_{\beta}
   + ({\cal L}_cx)x^{\star} + {1\over 2}[c,c]^{\alpha}c^{\star}_{\alpha}
   \Big)
\end{align}
In the expression for the action we integrate over a worldsheet. 
In the $h$-loop contribution the worldsheet is a surface of genus $h.$ 

Notice that one can introduce a notion of slightly degenerate complex structure assuming that the vectors $e$ and $\bar e$ can be linearly dependent on a family of closed curves on a worldsheet. (In a neighborhood of such a curve we should have a relation $\bar e=\lambda e +\rho f+...$ where tangent vectors $e$ and $f$ are linearly independent, $\rho=0$ is the equation of the curve and ... are higher order terms with respect to $\rho.$)
\section {String amplitudes}
\subsection { String amplitudes for critical string}

To represent the string theory in BV form we have applied  the general constructions of the 
Section \ref{sec:FromBRSTtoBV} to the action functional ${S_{\rm mat}[g,x]}.$ 
This functional depends on the metric $g_{\alpha\beta}$ on the worldsheet (on a compact surface of genus $h$) and a map ${\bf x}(\xi)=x^m(\xi)$ of this surface to $\mathbb{R}^d.$ This functional is invariant with respect to diffeomorphisms and Weyl transformations. We applied the standard BRST construction in this setting and  used (\ref {bv})  to get the BV action.
To describe string amplitudes we should add  marked points (punctures) $(\xi_1,..., \xi_n)$ on the worldsheet to this picture. Following \cite{Craps:2005wk} we will consider $\xi_i$ as dynamical variables on equal footing with the metric.

Using again the  constructions of the section \ref{sec:FromBRSTtoBV}  we get  the new BV action $S'_{\rm BV}$ with an extra term $c^{\alpha}(\xi_i)\xi_{\alpha }^{i\star}:$
\begin{equation}
\label{bbb}
S'_{\rm BV}=S_{\rm BV}+c^{\alpha}(\xi_i)\xi_{\alpha }^{i\star}
\end{equation}
where $S_{\rm BV}$ is defined by (\ref{BosonicMasterAction}).
As was noticed in section \ref{sec:FromBRSTtoBV} this functional obeys quantum master equation in the case when the volume is $Q$-invariant; this remark  forces us to use the diffeomorphism invariant measure  $\sqrt{g(\xi_1)}d^2\xi_1\cdots\sqrt{g(\xi_n)}d^2\xi_n$ on the space of marked points.

Let us consider  functionals $V_i (\xi_i)$  (vertices) which are invariant under diffeomorphisms.
 The typical examples  of such vertices are tachyoinic vertex $e^{i\bf{ px}(\xi)}$ and graviton vertex $ \epsilon_{kl}g^{\alpha\beta}\partial _{\alpha}  x^k(\xi)\partial_{\beta}x^l(\xi)e^{i{\bf px}(\xi)}.$
We can introduce a new action functional
\begin{equation}
\label{sh}
S''_{\rm BV}=S'_{\rm BV}+\sum \epsilon ^i V_i (\xi_i)
\end{equation} 
where $\epsilon _i$ are infinitesimally small. 

To define  string amplitudes it is convenient to work with BV-action functional that is obtained from (\ref {sh}) by means of ``integrating out'' Weyl ghosts. \footnote {If a solution $A$ of  the equation $\Delta A=0$ is defined on direct product of two  odd symplectic manifolds $\cal {Y}'$ and $\cal {Y}''$  we can obtain a solution of similar equation on $\cal {Y}'$ integrating over Lagrangian submanifolds $L\in \cal {Y}''.$ (See for example \cite {Cattaneo:2015vsa}.) In our case we integrate over Lagrangian submanifold   $\zeta^*=0$ of manifold  with coordinates $\zeta,\zeta^*.$} We obtain the new BV action ${\tilde S}_{\rm BV}$ given by the formula
$$e^{ {\tilde S}_{\rm BV}}=e^{S_{\rm mat}[g,x] \;+
\int \Big(
   ({\cal L}_cg)_{\alpha\beta}g^{\star\alpha\beta}
   + ((c^{\alpha}\partial_{\alpha})x^m)x_m^{\star} 
   + {1\over 2}[c,c]^{\alpha}c^{\star}_{\alpha} 
   - c^{\alpha}(\xi_i)\xi_{\alpha }^{i\star}+\sum \epsilon ^i V_i (\xi_i)
   \Big)}\delta ( g^{\star\alpha\beta}g_{\alpha\beta})
$$
Denoting the traceless part of $g^{\star\alpha\beta}$ by $b^{\alpha\beta}$ we can represent this action functional in the form
\begin{equation}\label{bv'}
{\tilde S}_{\rm BV}={\hat S}_{\rm BV}+\sum \epsilon ^i V_i (\xi_i)
\end{equation}
where
\begin{equation}
{\hat S}_{\rm BV}=S_{\rm mat}[g,x] \;+
\int \Big(
   ({\cal L}_cg)_{\alpha\beta}b^{\alpha\beta}
   + ((c^{\alpha}\partial_{\alpha})x^m)x_m^{\star} 
   + {1\over 2}[c,c]^{\alpha}c^{\star}_{\alpha}
   - c^{\alpha}(\xi_i)\xi_{\alpha}^{i\star}\Big)
\end{equation}
Now we can use the standard construction of the form $\Omega$ starting with the action functional 
${\tilde S}_{\rm BV}.$ However, we prefer to construct the form $\Omega$ starting with the functional ${\hat S}_{\rm BV}$ and including the factor $V_1...V_n$ into defining integral. (The form coming from the second construction can be obtained from the first one by means of differentiation with respect to parameters.)
 We consider a family of Lagrangian submanifolds parameterized by $ g_{\alpha\beta}^{(0)} \;,\xi^{(0)}_i\;$ taking
\begin{equation}
\label{gAlphaBeta0}
 g_{\alpha\beta}\;=\;  g_{\alpha\beta}^{(0)} \;,\;\xi_i=\xi^{(0)}_i\; ,\; x^{\star} = c^{\star} = 0
\end{equation}
The form $\Omega$ , restricted to one of these Lagrangian submanifolds looks  as follows: 
\begin{align}
&\Omega (g_{\alpha\beta}^{(0)} \;,\xi^{(0)}_i\;\delta g_{\alpha\beta}^{(0)} \;,d\xi^{(0)}_i\;)=
\nonumber\\[8pt]  
\;=\; &
\int [Dx\, Db\, D\xi^{*i}\,Dc] 
\sqrt{g(\xi^{(0)}_1)}V_1(\xi^{(0)}_1)\cdots\sqrt{g(\xi^{(0)}_n)}V_n(\xi^{(0)}_n) \;\times
\label{sca}\\  
& \times
   \exp\Big( S_{\rm mat} + 
   \int (\nabla_{\alpha}c_{\beta} + \nabla_{\beta}c_{\alpha})b^{\alpha \beta}\; + 
   \nonumber\\  
   & \phantom{\times\exp\Big( S_{\rm mat}} +\int \delta g^{(0)}_{\alpha\beta} b^{\alpha\beta} 
   +\xi^{*i}(c(\xi^i) - d\xi^{(0)}_i)\Big)  
\end{align}
Using this formula we can get an expression of $\Omega$  in terms of correlation functions of conformal field theory. This allows us to analyze the behavior of $\Omega$ with respect to Weyl transformations. It is easy to see that in our case of critical string this form is Weyl invariant if conformal fields  corresponding to vertices $V_i$ have conformal dimension $2$ (dimension $(1,1)$ in the language of complex geometry). In this case the form descends to the moduli space ${\cal M}_{h,n}$  of  compact complex curves of genus $h$ with $n$ marked points and to its Deligne-Mumford compactification $\bar {\cal {M}}_{h,n}$. Integrating over the fundamental cycle of $\bar {\cal {M}}_{h,n}$ we obtain the $h$-loop contribution to string amplitudes. To check this we notice that
after integration over $d\xi^{\star}$ (and omitting indices ${}^{(0)}$ for brevity) we get: 
\begin{align}
&
\int [Dx\, Db\, Dc] 
\Pi_j\left(
   \sqrt{g(\xi_j)}(-d\xi_j^1 + c^1(\xi_j))(-d\xi_j^2 + c^2(\xi_j))V_j(\xi_j)
\right)\;\times
\\  
& \times
   \exp\Big( 
   S_{\rm mat} + \int (\nabla_{\alpha}c_{\beta} + \nabla_{\beta}c_{\alpha})b^{\alpha \beta} 
   + \delta g_{\alpha\beta}b^{\alpha\beta}
   \Big)  
\label{StandardAmplitude}\end{align}
This result is equivalent to the standard expression for the string amplitude \cite{POS}. To see this we notice that
$\Pi_j(d\xi_j^1 + c^1(\xi_j))(d\xi_j^2 + c^2(\xi_j)) $ consist on $2^n$ summands; one of them gives the standard expression for string amplitudes with non-integrated vertices, another gives the standard expression with integrated vertices, and the rest correspond to the situation when some vertices are integrated and some are non-integrated. All these summands are equal, hence we obtain the standard answer up to a factor $2^n.$

Another way to calculate the string amplitudes is to work with infinitesimal deformations of BV 
action functional. Such deformations can be identified with (classical or quantum) observables. In string theory they can be considered as integrated vertices. Applying our approach to the deformation of BV action we obtain the standard expression of string amplitudes in terms of integrated vertices (see 
\cite {AM} for detail). 

An important method of calculation of scattering amplitudes in string theory is based on the consideration of {\em off-shell string amplitudes}. This is the best method  to calculate amplitudes when the mass gets quantum corrections.The  off-shell amplitudes  should be defined in such a way that the particle masses correspond to their poles  (in momentum representation)  and scattering amplitudes should be expressed in terms of residues in these poles.

To define off-shell string amplitudes for critical string one can consider  surfaces with marked points and local coordinate systems in the neighborhoods of these points 
\cite{Nelson:1988ic}, \cite {Sen:2014pia}. This is equivalent to consideration of surfaces with boundary. 
The BV formalism on manifolds with boundary was analyzed in \cite {Cattaneo:2015vsa}. It should be possible to 
\amklink{boundary/index.html}{combine our approach} with BV-BFV formalism of \cite  {Cattaneo:2015vsa}; 
these would lead to generalization of definitions given in  \cite{Nelson:1988ic}, \cite{Sen:2014pia}.

For non-critical strings very nice definition of off-shell amplitudes was suggested by 
A. Polyakov \cite{Polyakov:1987ez}; it works  well in our setting. 
Polyakov considers maps $\bf x (\xi)=x^m(\xi)$ of  a surface with marked points $\xi_1,...,\xi_k$ into $\mathbb{R}^d$ and includes the factor 
\begin{equation}
\label{fa}
\Pi _i\int\delta ({\bf x}_i-{\bf x}(\xi_i))\sqrt{g(\xi_i)}d^2\xi_i
\end{equation} 
in the functional integral that defines the partition function. Geometrically this means that we integrate 
over all surfaces in $\mathbb{R}^d$ that contain the points
${\bf x}_1, ..., {\bf x}_k\in \mathbb{R}^d $(surfaces  with pinned points $\{{\bf x}_i\}$ in Polyakov's terminology).  Doing the functional integral we obtain a function
$G({\bf x}_1,...,{\bf x}_k)$ that can be interpreted as off-shell amplitude in coordinate representation.  
The off- shell amplitude in the momentum 
representation $G({\bf p}_1,...,{\bf p}_k)$ can be defined as 
Fourier transform of $G({\bf x}_1,...,{\bf x}_k)$ or directly as a functional integral 
for partition function with insertion 
\begin{equation}
\label{ffa}
\Pi_j\int e^{i{\bf p}_j{\bf x}(\xi_j)}\sqrt{g(\xi_j)}d^2\xi_j
\end{equation}
Polyakov considers off-shell amplitudes only at tree level (genus zero surfaces), however they can be considered also in multi-loop case.

\section    {Pure spinor superstring}
We hope that  our  ideas  will lead to  better understanding of pure spinor formalism 
in superstring theory and to simplified expressions for amplitudes in this formalism. 

The worldsheet sigma-model of the pure spinor sigma-model has different versions
which are quasiisomorphic to each other, as usual in the topological field theory. 
There is a ``minimal version'', which (in case of Type II theory\footnote{For the heterotic string the right-moving variables are those
of the heterotic RNS formalism.}) describes matter fields 
$(x,\theta_L,\theta_R)$ and ``ghost fields'' $\lambda_L,\lambda_R$ constrained to live on the pure spinor cone:
\begin{equation} \label{ps}
(\lambda_L\Gamma^m\lambda_L) = (\lambda_R\Gamma^m\lambda_R) = 0
\end{equation}
The
\href{http://andreimikhailov.com/math/pure-spinor-formalism/flat-space/ActionAndBRST.html}{\textcolor{blue}{\bf flat
space sigma-model}} requires introduction of the momenta $p^L_+$  and $p^R_-$ conjugate to $\theta_L$ and $\theta_R$,
and the fermionic part of the action is of the first order in derivatives:
\begin{equation}
   \int d^2 z \left( p_+^L\partial_-\theta_L + p_-^R\partial_+ \theta_R \right)
\end{equation}
The action for pure spinors is, schematically:
\begin{equation}
\int d^2 z \;(w^L_{+}\partial_-\lambda_L + w^R_-\partial_+\lambda_R)
\end{equation}
where the ``conjugate momenta'' $w^L_{+},w^R_-$ take values in the cotangent bundle of the pure 
spinor cone.  
The bosonic part of the action is the usual $\int d^2z \;\partial_+ x^m\partial_- x^m$.

The model is invariant under a fermionic nilpotent symmetry $Q$.
Importantly, it splits (for Type II case) into the sum of left and right symmetries:
\begin{equation}
   Q = Q_L + Q_R
\end{equation}
such that the conserved currents corresponding to $Q_L$ and $Q_R$ are holomorphic and
antiholomorphic, respectively.

In the case of flat target space, it is easy to obtain the corresponding BV action functional: 
for every field $\Phi$ one should add its antifield $\Phi^*$  and a   term in the action having the 
form $(Q \Phi) \cdot \Phi^*.$ (This is a special case of general construction described in Sec 5; see (\ref {bv}).) 

However, the solution of Eq. (\ref {ph} ) requires different methods. As a  first step, let us 
restrict ourselves to the left sector\footnote{applying this to the ``full'' sigma-model, {\it i.e.} left plus right sector, is work in progress in collaboration with R.~Lipinski~Jusinskas}. 
The explicit form of Eq. (\ref{ph}) for the left sector of the pure spinor string is:
\begin{align}
   & \{S_{BV}, a(\xi)\} + {1\over 2} \{a(\xi),a(\xi)\} = {\cal H}\langle\xi\rangle
   \label{LeftSectorPureSpinor}\\  
   \mbox{\tt\small where } 
   & {\cal H}\langle\xi\rangle \;=\; (\xi^z\partial_z x^m) x^{\star}_m + (\xi^z\partial_z\theta)\theta^{\star} + ({\cal L}_{\xi} p_+)p^{\star +} \;+
   \nonumber\\   
   & \quad\quad\quad + (\xi^z\partial_z\lambda_L)\lambda_L^{\star} + ({\cal L}_{\xi} w_+) w^{\star +} 
\end{align}
--- this has to be solved for the unknown $a(\xi)$; notice that ${\cal H}\langle\xi\rangle$ is
linear in $\xi$, but $a(\xi)$ does not have to be linear in $\xi$. 
(We have assumed that $\Delta a(\xi)=0$ ; otherwise we should add an ill-defined term $\Delta a(\xi)$.)
One solution can be obtained as follows.
Since the worldsheet theory is conformal, a holomorphic vector field $\xi^+$ is
a symmetry; it is generated by $\xi^+T_{++}$.  It was shown by Berkovits that the energy-momentum
tensor is BRST-trivial: $T_{++} = Q_Lb_{++}$ (even off-shell) where $b_{++}$ is a composite $b$-ghost.
This means that one should expect that the worldsheet action can be included 
into topological conformal field theory. A rigorous proof of this statement is still unknown;
the most convincing treatment of this problem was given in 
\cite {Hoogeveen:2007tu}\footnote{that paper contains also the  
calculation of superstring amplitudes in the framework of BV-formalism; some ideas of this 
calculation can be used in our approach.} 
Notice that  $\xi^+b_{++}$ is a holomorphic current and therefore also corresponds to
some symmetry. We can identify $a(\xi) = \Phi\langle\xi\rangle$, a BV Hamiltonian generating the infinitesimal action 
of that symmetry. Then  the second term in (\ref{LeftSectorPureSpinor}) vanishes and this equation is satisfied. However we hope that there exist simpler solutions of Eq. (\ref{LeftSectorPureSpinor})  with non-vanishing second term; we leave this
question for future work. We believe, that applying the techniques described 
above one can not only justify the pure spinor formalism, but also simplify the formulas 
(hopefully we can avoid using the complicated and not very well defined "composite $b$-ghost").
 
\appendix

\section{Some useful formulas}
\amklink{BV-formalism/index.html}{BV phase space} is an odd symplectic supermanifold $M$
with a nondegenerate closed odd 2-form $\omega$. For any $F\in \mbox{Fun}(M)$ we can define its Hamiltonian
vector field. We will think of this vector field as a first order linear differential operator,
acting on $\mbox{Fun}(M)$:
\begin{equation}
G \;\mapsto\; \{F,G\}=F\stackrel{\leftarrow}{\partial\over\partial Z^A} \pi^{AB}(Z) {\partial\over\partial Z^B} G\end{equation}
and denote this operator $\{F,\_\}$. (Here $\pi^{AB}(Z)$ is a matrix inverse to $\omega _{AB}(Z)$.) By definition:
\begin{equation}
dF\;=\;(-)^{\overline{F} + 1} \iota_{\{F,\_\}} \omega
\end{equation}
where $\iota$ is the operator of contraction, satisfying $[\iota_V,d]={\cal L}_V$. This implies:
\begin{equation}\label{PBFG}
\{F,G\} = \iota_{\{F,\_\}} dG = (-)^{\bar{G}+1}\iota_{\{F,\_\}} \iota_{\{G,\_\}} \omega
\end{equation}
In coordinates:  
\begin{align}
\omega\;=\; & \underline{dZ}^A \underline{dZ}^B\omega_{AB} 
\\   
\omega_{AB} \;=\; & (-1)^{(\bar{A}+1)(\bar{B}+1)}\omega_{BA}
\\    
d \;=\; & \underline{dZ}^A {\partial\over\partial Z^A}
\\    
\iota_{V}\;=\;& V^A{\partial\over\partial \underline{dZ}^A}
\\      
\pi^{AB}\;=\; &
(-1)^{1 + (\bar{A}+1)(\bar{B}+1)}\pi^{BA}
\end{align}
Locally it is possible to choose the Darboux coordinates: 
\begin{align}
\{F,G\}\;=\; &
F\left(
  \stackrel{\leftarrow}{\partial\over\partial \phi^{\star}_A}\;{\partial\over\partial\phi^A} \;-\;
  \stackrel{\leftarrow}{\partial\over\partial \phi^A}\;{\partial\over\partial\phi^{\star}_A} 
\right)G 
\label{DarbouxCoordinates}
\\       
\omega \;=\; &
 (-1)^A \underline{d\phi}^A \underline{d\phi}^{\star}_A
\end{align}

If the manifold  $M$ is equipped with a volume element  (with a density)  we can define the odd Laplacian acing on functions by the formula
\begin{equation}
\label{ }
\Delta F=div \{F,\_\}
\end{equation}
where $div$ stands for the divergence of vector field with respect to the volume element.

The volume element should be chosen in such a way that $\Delta^2=0.$
The relation between 
\amklink{BV-formalism/OddLaplace.html}{odd Laplace operator} 
and $\{\_,\_\}$ is: 
\begin{align}
\Delta(XY) \;=\;& (\Delta X)Y + (-)^{\bar{X}}\Delta Y + (-)^{\bar{X}}\{X,Y\}
\\ 
\Delta e^{\Phi}\;=\;& \left(\Delta \Phi + {1\over 2}\{\Phi,\Phi\}\right) e^{\Phi}
\label{DeltaEPhi}
\end{align}
In Darboux coordinates $\Delta$ is:
\begin{equation}\label{DeltaInDarboux}
\Delta \;=\; (-1)^{\bar{A} + 1}  {\partial\over\partial \phi^{\star}_A}{\partial\over\partial \phi^A}
\end{equation}
One can prove that  $\Delta X$ given by this formula does not depend on the choice of Darboux coordinates if $X$ transforms as a semidensity
 (recall that semi-densities transform as square roots of densities= volume elements). Hence for any odd symplectic manifold  one can define $\Delta$ on semi-densities (volume element is not necessary), see \cite{Khudaverdian:1999}.

\section{Definition of $\Omega$ using marked points}\label{sec:UsingMarkedPoints}
Let ${\rm LAG}_+$ denote the space of Lagrangian submanifolds with marked points. A point of $\rm LAG_+$ 
is a pair $(L,a)$ where $L\in \rm LAG$ and $a\in L$. This defines the double fibration:
\begin{equation}
   M \stackrel{p}{\longleftarrow} {\rm LAG}_+ \stackrel{\pi}{\longrightarrow} {\rm LAG}
\end{equation}
Given  $v\in \Pi T_{(L,a)}{\rm LAG}_+$, we can consider two projections $\pi_*v\in \Pi T_L{\rm LAG}$
and $p_*v\in \Pi T_aM$. We will define $\Omega$ is a pseudo-differential form, {\it i.e.} a function of
$L,a,v$. It will depend on $v$ only through $\pi_*v$. 
We can characterize $\pi_* v$ as a section of $\Pi TM|_L$ modulo $\Pi TL$. We then define $\sigma$ as
follows:
\begin{align}
   \sigma \;\in\; & \mbox{Fun}(L)
   \nonumber \\    
   d\sigma \;=\; & - \left.\left(\iota_{\pi_*v} \omega\right)\right|_L
   \label{DefDSigma}
   \\  
   \sigma(a) \;=\; & 0
   \label{SigmaVanishiesAtMarkedPoint}
\end{align}
This definition specifies $\sigma$ as a linear function of $v$ , {\it i.e.} as a one-form on $ {\rm LAG}_+$
In order to make sense of $\iota_{\pi_*v}\omega$ we must think of $\pi_*v$ as a section of $\Pi TM$; the fact 
that it is only defined up to tangent to $TL$ does not matter because $L$ is isotropic. Eq. (\ref{SigmaVanishiesAtMarkedPoint}) 
eliminates the ambiguity, and we can now safely define a function $\Omega$ on $ \Pi T{\rm LAG}_+$ (a pseudodifferential form on $ {\rm LAG}_+$)
as in Eq. (\ref{omega-in-BV-formalism}):
\begin{equation}
\Omega (L,a,v)\;=\; \int_L e^{S_{\rm BV} + \sigma}
\end{equation}
More generally, for every function $F$ on $M$ we define:
\begin{equation}
\label {FFF}
   \Omega\langle F\rangle(L,a,v)\;=\;\int_LFe^{S_{\rm BV}+\sigma}
\end{equation}
We will now prove the following formula:
\begin{equation}
\label{OmegaWithF}
(d - p^*\omega)\;\Omega \langle F\rangle 
\;=\; 
- \Omega \left\langle \Delta  F+ \{S_{\rm BV}\,,\,F\}\right\rangle 
\end{equation}
\paragraph     {Comment}
As a straightforward generalization, we can consider a product of $\Omega$ with the pullback
under $p$ of any differential or pseudo-differential form $\nu$ on $M$.
\amklink{omega/Descent\_To\_LAGs.html\#(part.\_.Upgrade\_\_to\_)}{It satisfies:} 
\begin{equation}\label{SeveraDifferential}
d\left(p^*\nu\;\Omega \langle F\rangle\right) \;=\; 
(-)^{|\nu|+1} p^*\nu\;\Omega \left\langle \Delta  F+ \{S_{\rm BV}\,,\,F\}\right\rangle 
\;+\;p^*\left(d\nu + \omega\nu\right)\;\Omega\left\langle F\right\rangle
\end{equation}
Notice the appearance of the nilpotent operator $d + \omega$ which was studied in \cite{Severa2006}.

\paragraph     {Proof}
We take a family of Lagrangian submanifolds with
marked points $(L(\lambda),a(\lambda))$ and represent it in the form

\begin{align}
   L(\lambda) \;=\; & g(\lambda)L_0
   \\  
   a(\lambda) \;=\; & g(\lambda)a_0
\end{align}
where $g(\lambda)$ are volume preserving canonical transformations
(locally this is always possible).

It is sufficient to analyze the restriction $ \Omega\langle F\rangle(\lambda, d\lambda)$ of the form (\ref {FFF}) to this family.

As in Section 3 using the canonical transformations $g(\lambda)$  we can construct a family of action functionals $S_{\lambda}$ and corresponding forms that will be denoted by $\tilde \Omega$ and ${\tilde \Omega}\langle F\rangle.$  These forms do not coincide with the forms $ \Omega\langle F\rangle(\lambda, d\lambda)$   constructed by means of family of Lagrangian submanifolds with marked points, but they are closely related.
As we noticed in Section 3 the second summand in the exponential in the formula defining $ {\tilde \Omega}\langle F\rangle(\lambda, d\lambda)$  is the Hamiltonian of the infinitesimal canonical transformation governing the variation of $S_{\lambda}$. The second summand in the formula defining  $ \Omega\langle F\rangle(\lambda, d\lambda)$ is the Hamiltonian
${\cal H}(\lambda,d\lambda)$ of the infinitesimal canonical transformation\footnote{It is related to the $B_a$ used in Section \ref{sec:FamiliesOfActionFunctionals} as follows:
${\cal H}(\lambda,d\lambda)(g(\lambda)x) = \sum d\lambda^a B_a(\lambda, x)$.} governing the variation of $L_{\lambda}$. 
They coincide up to a constant summand. This constant can be calculated from (\ref {DefDSigma}). 
\amklink{omega/Descent\_To\_LAGs.html\#(part.\_.Upgrade\_\_to\_)}{We obtain} 
\begin{equation}
\label{ }
 \Omega\langle F\rangle(\lambda, d\lambda)=C{\tilde \Omega}\langle F\rangle(\lambda, d\lambda)
\end{equation}
where $C=  e^{-{\cal H}(\lambda,d\lambda)(g(\lambda)a_0)}.$ (One can say that $C$ is expressed in terms of the value of the Hamiltonian of the infinitesimal canonical transformation at the marked point.) 

We have calculated already the differential of ${\tilde \Omega}\langle F\rangle(\lambda, d\lambda)$. But we also have to evaluate $d_{\Lambda}$ of the prefactor $C$. Using Eq. (\ref{PBFG}), Appendix, and $p^*\omega = {1\over 2}\left(\iota\left(d\lambda^k{\partial a^A\over\partial\lambda^k}{\partial\over\partial a^A}\right)\right)^2\omega$ we get:
\begin{align}
   & d_{\Lambda} e^{-{\cal H}(\lambda,d\lambda)(ga_0)}\;=\; 
   \\    
   \;=\; &
   e^{-{\cal H}(\lambda,d\lambda)(ga_0)}\left(
      - (d_{\Lambda} {\cal H}(\lambda, d\lambda)) (ga_0) 
      - \{{\cal H}(\lambda,d\lambda)\,,\,{\cal H}(\lambda,d\lambda)\}(ga_0)
   \right)\;=\;
   \\   
   \;=\; &
   - {1\over 2}e^{-{\cal H}(\lambda,d\lambda)(ga_0)}
       \{{\cal H}(\lambda,d\lambda)\,,\,{\cal H}(\lambda,d\lambda)\}(ga_0)\;=\;
       \\  
       \;=\;& 
       {1\over 2} e^{-{\cal H}(\lambda,d\lambda)(ga_0)}
       ((\iota_{\{{\cal H}(\lambda,d\lambda)\,,\,\_\}})^2\omega )(ga_0) \;=\;
       \\    
       \;=\;&
       {1\over 2} e^{-{\cal H}(\lambda,d\lambda)(ga_0)}
       (\iota_{\{{\cal H}(\lambda,d\lambda)\,,\,a^A\}\partial/\partial a^A})^2\omega(a)|_{a=ga_0} \;=\;
       \\ 
       \;=\;&
       {1\over 2} e^{-{\cal H}(\lambda,d\lambda)(ga_0)}
       (\iota_{d\lambda^k(\partial_ka^A)\partial/\partial a^A})^2\omega(a)|_{a=ga_0} \;=\;
       e^{-{\cal H}(\lambda,d\lambda)(ga_0)}p^*\omega
\end{align}
This concludes the proof.

Given a ``symplectic potential'' $\alpha$ satisfying $d\alpha = \omega$ we can construct a closed form as follows:
\begin{align}\label{ESymplecticPotentialOmega}
   \Omega_+ = (p^*e^{-\alpha})\;\int_L e^{\sigma}
\end{align}

We will choose the following ansatz for the equivariantly closed analogue of $\Omega$:
\begin{equation}\label{AnsatzForEquivariantOmegaPlus}
   \Omega^{\tt C}_+ \;=\; (p^*\nu)\;\int_L e^{S + \sigma + \Phi(h)}
\end{equation}
where $\nu$ is of the same formal type as a Cartan cochain:
\begin{align}\label{CartanComplexOfM}
   \nu \;\in\; \mbox{Fun}\left((\Pi T M)\times {\bf h}\right)
\end{align}
The expression defined in Eq. (\ref{AnsatzForEquivariantOmegaPlus}) is a cocycle of the Cartan complex of equivariant cohomology 
of ${\rm LAG}_+$ if in addition to (\ref {ph}) we have \begin{align}
  \left(d + \omega - \iota_{\{h,\_\}} + h\right)\nu \;=\;0
\label{NuEquivariantlyClosed}
\end{align}

Even though $\nu$ lives in the same space as cochains of the Cartan complex, the
differential defined by Eq. (\ref{NuEquivariantlyClosed}) is different. (The Cartan differential would be $d - \iota_{\{h,\_\}}$.)

\paragraph     {Comment} In particular, {\em when} we can choose an $H$-invariant
``symplectic potential'' $\alpha$ such that $d\alpha = \omega$, Eq. (\ref{NuEquivariantlyClosed}) has a simple solution:
\begin{equation}
   \nu = e^{\alpha}
\end{equation}
\paragraph     {Proof of $\Omega_+^{\tt C}$ being equivariantly closed}
We have to prove that:
\begin{equation}
   \left(d - \iota_{\{h,\_\}}\right)\Omega^{\tt C}_+\;=\;0
\end{equation}
where $d$ is the de Rham differential on ${\rm LAG}_+$. The action of $d$ is given by Eq. (\ref{SeveraDifferential}). 
The action of $\iota_{\{h,\_\}}$ on $\sigma$ is essentially as in Eq. (\ref{ActionOfIotaOnSigma}), but we have to remember to subtract
the compensating constant to make sure that $\sigma$ vanishes at the marked point; therefore:
\begin{equation}
   \iota_{\{h,\_\}}\sigma = h - h(a)
\end{equation}
The vanishing of $(d-\iota_{\{h,\_\}})\Omega^{\tt C}_+$ when Eqs. (\ref{ph}) and (\ref{NuEquivariantlyClosed}) are satisfied follows from direct 
computation.

\section{Central extension of the group of canonical transformations}\label{sec:CentralExtension}
In this Section we will give a precise definition of $\Omega$ using a well-defined closed PDF $\widehat{\Omega}$ on a central extension $\widehat{G}$ of the group 
of  canonical transformations.\footnote{The existence of  a central extension of the group of canonical transformations (symplectomorphisms) of odd symplectic manifold $M$ can be proven in the same way as for an even symplectic manifold. Namely, as in the even case one constructs  a bundle with connection over $M$, the fiber of this bundle is a one-dimensional odd vector space. The group $\widehat{G}$ can be defined as a group of transformations of the total space of the bundle that are compatible with the fibration (transform fibers into fibers), induce canonical transformation on the base and are compatible with connection.}  This group  is infinite-dimensional, however, in this section we will keep the notation $d$ for the de Rham differential on the group and on the space of Lagrangian submanifolds LAG.

\subsection{Definition of $\widehat{\Omega}$}
Let us consider the Lie superalgebra $\Pi \mbox{Fun}(M)$ with the commutator given by the
odd Poisson bracket. It is a central extension of the Lie superalgebra of Hamiltonian vector
fields which we denote $\bf g$; therefore we  denote it $\widehat{\bf g}$:
\begin{equation}
\widehat{\bf g} \;=\; \Pi\mbox{Fun}(M)
\end{equation}
We consider the central extension of the
group of canonical transformations $\widehat{G}$, whose Lie algebra is  $\widehat{\bf g}$.

As a variation on our theme, we will now construct a map from $\rm LAG$ to the space of closed 
PDFs on $\widehat{G}$, which we will call $\widehat{\Omega}$:
\begin{align}
\widehat{\Omega}\;\in\; & \mbox{Fun}({\rm LAG}\times \Pi T\widehat{G})
\\    
\widehat{\Omega}(L,\widehat{g},d\widehat{g})\;=\;&
\int_{{g}L} \exp\left(S_{\rm BV} + d\widehat{g}\widehat{g}^{-1}\right)
\label{DefOmegaHat}
\end{align}
Here following \cite{Khudaverdian:1999} we consider $\exp(S_{\rm BV})$ as a semidensity,
 $d\widehat{g}\widehat{g}^{-1}$ is the right-invariant form on $\widehat{G}$ taking values in the Lie algebra (Maurer-Cartan form),and $g$ stands for an element of $G$ corresponding to $\widehat{g}\in \widehat{G}$. In Eq. (\ref{DefOmegaHat}) we consider $d\widehat{g}\widehat{g}^{-1}$ as a 
function on $M$, using the fact that the Lie algebra of $\widehat{G}$ is $\Pi \mbox{Fun}(M)$.  This form  satisfies
the Maurer-Cartan equation:
\begin{equation}\label{MaurerCartanForGHat}
d(d\widehat{g}\widehat{g}^{-1}) + 
{1\over 2} \{d\widehat{g}\widehat{g}^{-1}\,,\,d\widehat{g}\widehat{g}^{-1}\}\;=\;0
\end{equation}
This $\widehat{\Omega}$ is closed as a PDF on $\widehat{G}$, {\it i.e.}:
\begin{align}
d\widehat{\Omega}\;=\;& 0 \label{HatOmegaIsClosed}
\\    
\mbox{\tt\small where } d \;=\; & d\widehat{g}{\partial\over\partial\widehat{g}}
\end{align}
The proof of Eq. (\ref{HatOmegaIsClosed}) is a \amklink{omega/Definition.html}{straightforward computation} very similar to the computations in Section \ref{sec:FamiliesOfActionFunctionals}.  

\vspace{10pt}
\noindent
We must stress that this $\widehat{\Omega}$ is well-defined (does not contain any ambiguities). 

\subsection{How to build a form on $\rm LAG$ starting from $\widehat{\Omega}$}
Since $G$ (and therefore $\widehat{G}$) acts on $\rm LAG$, there is a natural projection:
\begin{equation}\label{PiHat}
\widehat{\pi} \;:\;{\rm LAG}\times \Pi T\widehat{G}\to \Pi T {\rm LAG}
\end{equation}
However, it is {\em not true} that  $\widehat{\Omega}$ is constant along the fibers of $\widehat{\pi}$. Indeed,
for a $\xi\in  \mbox{Lie}(\mbox{St}(L_0))$, where $\mbox{St}(L_0)$ stands for the stable subgroup of $L_0\in {\rm LAG}$ in $\widehat G$ one can check that  the restriction of $\xi$ on $L_0$ is a constant $c$. 
Using $\widehat{g}\widehat{\xi}\widehat{g}^{-1} = \widehat{\xi}\circ g^{-1}$ we get:
\begin{equation}
\widehat{\Omega}(L_0,\, \widehat{g},\, d\widehat{g} + \widehat{g}\widehat{\xi})\;=\;
k\widehat{\Omega}(L_0,\widehat{g},d\widehat{g})
\end{equation}
where $k$ is some number. Therefore $\widehat{\Omega}$ does not automatically provide  a PDF on $\rm LAG$.

We could impose some additional restrictions, such as ghost number symmetry\footnote{We do not require that the semidensity be invariant
under the ghost number symmetry; just that $d\widehat{g}\widehat{g}^{-1}$ have ghost number $-1$},
which would guarantee that $k=0$. 

Let us suppose now that a subset of ${\rm LAG}$ is represented in the form $g(\lambda)L_0$ where $g(\lambda)\in G, \lambda\in\Lambda.$ Assume that  we can find a ``lift'' $\widehat{g}(\lambda)$ of $g(\lambda)$ to $\widehat{G}$. 
Then we can define a closed form
\begin{equation}
\Omega(L,dL) \;=\; \widehat{\Omega}(L_0, \widehat{g}(\lambda), d(\widehat{g}(\lambda)))
\end{equation}
  
This coincides with the ``tentative'' definition of Section \ref{sec:FamiliesOfLAG}, because the restriction of 
$d\hat{g}\hat{g}^{-1}$ to ${g}L_0$ gives $\sigma$. This is a general fact, true both in classical mechanics and 
in BV formalism. In classical mechanics it is essentially the Hamilton-Jacobi equation, 
which describes the evolution of a Lagrangian submanifold (specified by a generating function 
usually called $S$) under the Hamiltonian flow. It says that ${\partial S\over\partial t}$ equals the restriction of $H$ 
on $L$ plus a constant (which can depend on $t$).

Notice that by the variation of   $\widehat{g}(\lambda)$ the form $\Omega(L,dL)$ obviously remains in the same cohomology class.

\section*{Acknowledgments}
We are grateful to Nathan~Berkovits, Alberto~Cattaneo, Alexei~Kotov, Misha~Movshev, 
John~Murray, Pavel~Mnev, Sasha~Polyakov and Kostas~Skenderis for useful discussions.
The work of A.M. was partially supported by the FAPESP grant 2014/18634-9 
``Dualidade Gravitac$\!\!,\tilde{\rm a}$o/Teoria de Gauge'',
and in part by the RFBR grant 15-01-99504 ``String theory and integrable systems''.


\begin{thebibliography}{10}

\bibitem {ST} A.~ Schwarz {\it {Topological quantum field theories}} Proceedings of ICMP (2000)
\bibitem {SC} A.~Schwarz. {\it Supergravity, complex geometry and G-structures}. Communications in Mathematical Physics. 1982 Mar 1;87(1):37-63.
\bibitem {SD}       A.~      Schwarz . {\it On the definition of superspace}. Teoreticheskaya i Matematicheskaya Fizika. 1984;60(1):37-42. 
English version: 
Theoretical and Mathematical Physics, 1984, 60:1, 657?660

\bibitem{Schwarz:1992nx}
A.~Schwarz, {\it {Geometry of Batalin-Vilkovisky quantization}},  {\em
  Commun.Math.Phys.} {\bf 155} (1993) 249--260 doi: {\bf 10.1007/BF02097392}
  [{\tt arXiv/hep-th/9205088}].
\bibitem{Schwarz:1992gs}
A. Schwarz, {\it {Semiclassical approximation in Batalin-Vilkovisky
  formalism}},  {\em Commun.Math.Phys.} {\bf 158} (1993) 373--396 doi: {\bf
  10.1007/BF02108080} [{\tt arXiv/hep-th/9210115}].

\bibitem{Khudaverdian:1999}
O.~M. Khudaverdian, {\it {Delta-Operator on Semidensities and Integral
  Invariants in the Batalin-Vilkovisky Geometry}}, {\tt arXiv/9909117 }.

\bibitem{Schwarz:1993xp}
A.~ Schwarz, {\it {Symmetry transformations in Batalin-Vilkovisky
  formalism}},  {\em Lett.Math.Phys.} {\bf 31} (1994) 299--302 doi: {\bf
  10.1007/BF00762792} [{\tt arXiv/hep-th/9310124}].
   \bibitem{Meinrenken}
E.~Meinrenken, {\it Equivariant cohomology and the cartan model}, {\tt
  http://www.math.toronto.edu/mein/research/enc.pdf }.
\bibitem {COS}K.~Costello, 2011. {\it Renormalization and effective field theory} (Vol. 170). Providence: American Mathematical Society.
\bibitem{Cattaneo:2015vsa}
A.~S. Cattaneo, P.~Mnev, and N.~Reshetikhin, {\it {Perturbative quantum gauge
  theories on manifolds with boundary}}, {\tt arXiv/1507.01221 }.
\bibitem{SCHP} A.~Schwarz, (1979). {\it The partition function of a degenerate functional. }Communications in Mathematical Physics, 67(1), 1-16.
  \bibitem{KON}M. Kontsevich, {\it Feynman diagrams and low-dimensional topology.} InFirst European Congress of Mathematics Paris, July 6?10, 1992 1994 (pp. 97-121). Birkhäuser Basel.

\bibitem{WIT}E. Witten , {\it Chern-Simons gauge theory as a string theory.} InThe Floer memorial volume 1995 (pp. 637-678). Birkhäuser Basel.
\bibitem{Craps:2005wk}
B.~Craps and K.~Skenderis, {\it {Comments on BRST quantization of strings}},
  {\em JHEP} {\bf 0505} (2005) 001 doi: {\bf 10.1088/1126-6708/2005/05/001}
  [{\tt arXiv/hep-th/0503038}].
  \bibitem {POS} J.~Polchinski, {\it String Theory} Cambridge Univ. Press
\bibitem {AM} A.~Mikhailov (in preparation)

\bibitem{Nelson:1988ic}
P.~C. Nelson, {\it {Covariant insertion of general vertex operators}},  {\em
  Phys.Rev.Lett.} {\bf 62} (1989) 993 doi: {\bf 10.1103/PhysRevLett.62.993}.

\bibitem{Sen:2014pia}
A.~Sen, {\it {Off-shell Amplitudes in Superstring Theory}},  {\em Fortsch.
  Phys.} {\bf 63} (2015) 149--188 doi: {\bf 10.1002/prop.201500002} [{\tt
  arXiv/1408.0571}].

\bibitem{Polyakov:1987ez}
A.~M. Polyakov, {\it {Gauge Fields and Strings}},  {\em Contemp. Concepts
  Phys.} {\bf 3} (1987) 1--301.

\bibitem{Hoogeveen:2007tu}
J.~Hoogeveen and K.~Skenderis, {\it {BRST quantization of the pure spinor
  superstring}},  {\em JHEP} {\bf 11} (2007) 081 doi: {\bf
  10.1088/1126-6708/2007/11/081} [{\tt arXiv/0710.2598}].

\bibitem{Severa2006}
P.~{\v{S}}evera, {\it On the origin of the bv operator on odd symplectic
  supermanifolds},  {\em Letters in Mathematical Physics} {\bf 78} (2006),
  no.~1 55--59 doi: {\bf 10.1007/s11005-006-0097-z} url: {\bf
  http://dx.doi.org/10.1007/s11005-006-0097-z}.


\bibitem {ZZ} A. Zamolodchikov, Al. Zamolodchikov, http://qft.itp.ac.ru/ZZ.pdf 


\end{thebibliography}

\def\cprime{$'$} \def\cprime{$'$}
\providecommand{\href}[2]{#2}\begingroup\raggedright\endgroup

\end{document}